\pdfoutput=1

\documentclass{raa}
\usepackage{mathptmx}
\usepackage{ae,aecompl}

\usepackage{graphicx}	
\usepackage{amsmath}	
\usepackage{amssymb}	
\usepackage{pifont}
\usepackage{txfonts}
\usepackage{url}
\usepackage{subfigure}
\usepackage{longtable}
\usepackage{lscape}
\usepackage{multirow}
\usepackage{color}
\usepackage{textcomp}
\usepackage{natbib}
\usepackage[colorlinks,linkcolor=red,anchorcolor=green,citecolor=blue]{hyperref}

\newcommand{\HI}{H\,{\small{I}} }

\begin{document}

\title{Performance of FAST with an Ultra-Wide Bandwidth Receiver at 500-3300\,MHz}


   \volnopage{Vol.0 (2023) No.0, 000--000}      
   \setcounter{page}{1}          
   \author{Chuan-Peng Zhang\inst{1,2},
      Peng Jiang\inst{1,2},
      Ming Zhu\inst{1,2},
      Jun Pan\inst{1,3},
      Cheng Cheng\inst{1},
      Hong-Fei Liu\inst{1,2},
      Yan Zhu\inst{1,2},
      Chun Sun\inst{1,2},
      FAST Collaboration\inst{1}
      }

   \institute{National Astronomical Observatories, Chinese Academy of Sciences, Beijing 100101, China; {\it cpzhang@nao.cas.cn} \\
    \and
    Guizhou Radio Astronomical Observatory, Guizhou University, Guiyang 550000, China\\
    \and 
    College of Earth Sciences, Guilin University of Technology, Guilin 541004, China\\
   }


\abstract{The Five-hundred-meter Aperture Spherical radio Telescope (FAST) has been running for several years. A new Ultra-Wide Bandwidth (UWB) receiver, simultaneously covering 500-3300\,MHz, has been mounted in the FAST feed cabin and passed a series of observational tests. The whole UWB band is separated into four independent bands. Each band has 1048576 channels in total, resulted in a spectral resolution of 1\,kHz. At 500-3300\,MHz, the antenna gain is around 14.3-7.7 K\,Jy$^{-1}$, the aperture efficiency is around 0.56-0.30, the system temperature is around 88-130\,K, and the HPBW is around 7.6-1.6\,arcmin. The measured standard deviation of pointing accuracy is better than $\sim$7.9\,arcsec, when zenith angle (ZA) is within 26.4$^\circ$. The sensitivity and stability of the UWB receiver are confirmed to satisfy expectation by spectral observations, e.g., \HI and OH. The FAST UWB receiver already has a good performance for taking sensitive observations in various scientific goals. 
 }

   \authorrunning{Chuan-Peng Zhang, et al.}                     
   \titlerunning{FAST UWB Receiver at 500-3300\,MHz}       

   \maketitle


\section{Introduction}    
\label{sect:intro}

The Five-hundred-meter Aperture Spherical radio Telescope (FAST) with an effective diameter of 300\,m has obtained many groundbreaking achievements, for example in observations of pulsar, fast radio burst, star formation, galaxy evolution \citep[e.g.,][]{Cheng2020,Han2021,Li2021,Niu2022,Ching2022,Xu2022}, since FAST began its commission when the construction was completed on September 25, 2016 \citep{Nan2011,Jiang2019,Jiang2020}. Till now, FAST mainly worked at frequencies of 1000-1500\,MHz with a 19-beam receiver. Recently, a new cryogenic UWB receiver at 500-3300\,MHz has been developed by \citet{Liu2022} and mounted in the FAST feed cabin for science observations. In view of the 19-beam receiver occupies all three Hellium cryogenic compressors and most of the space of the feed cabin, currently it has no enough space for the UWB receiver to place any more cryogenic compressor. Now, the FAST UWB receiver has passed a series of test work, and could be carried out kinds of spectral observations.



At 500-3300\,MHz, the FAST UWB receiver is ideally able to simultaneously cover 330 radio combination lines (RRLs) for H$n\alpha$, He$n\alpha$, and C$n\alpha$ ($n=235-126$), respectively. This could help us to investigate the active star formation regions in the Milky Way \citep[e.g.,][]{Chen2020,Zhang2021,Hou2022}. Furthermore, the UWB receiver could simultaneously cover the Hydrogen (\HI at 1420.406\,MHz) and Hydroxyl radical (OH at 1612.231, 1665.402, 1667.359, and 1720.530\,MHz) lines, and also their high redshift signals with $z\lesssim1.8$. This gives us an opportunity to study the star formation and evolution not only in the Milky way, but also in the nearby galaxies especially to provide us with multiwavelength spectral data. In addition, the UWB receiver has been able to catch the Methyladyne (CH) line at 3263.794\,MHz. This would return us a high spatial resolution data ($\sim$1.6$'$) for better inspecting our Galaxy. Furthermore, the UWB receiver has a sufficient sensitivity and high enough spectral resolution (1\,kHz). This allows us to well study the kinematic information of star formation in the Milky Way and the hyperfine structures of some spectral lines (e.g., OH at $\sim$1665.402\,MHz).

Thanks to the advantageous characteristics of the FAST, we are able to complete a series of observational tests in a short time. In this report, we mainly present the performance of the FAST UWB receiver and relevant antenna parameters at 500-3300\,MHz. General parameters of FAST UWB receiver are listed in Table\,\ref{tab_uwb_para}. In Section\ref{sec:uwb_para}, we introduce the measurement parameters of the UWB receiver system including the noise dipole, beam properties, pointing accuracy, antenna gain, aperture efficiency, and system temperature. In Section\,\ref{sec:backend_obs}, we present the properties of the spectral backend and the measurement results in spectral \HI and OH observations. Summary is presented in Section\,\ref{sect:summary}.

\begin{table*}
\caption{General parameters of FAST with UWB receiver.}
\label{tab_uwb_para} 
\centering \small  
\setlength{\tabcolsep}{1.6mm}{
\begin{tabular}{l|cccc}
\hline \hline
 FAST UWB receiver & UWB-1 & UWB-2 & UWB-3 & UWB-4 \\
\hline
Total frequency range (MHz) & $0\sim1100$ & $800\sim1900$ & $1600\sim2700$ & $2400\sim3500$ \\
Total channel number & 1048576 & 1048576 & 1048576 & 1048576 \\
Spectral resolution (Hz) & 1049.04 & 1049.04 & 1049.04 & 1049.04 \\
Effective frequency range (MHz) & $500\sim1000$ & $900\sim1800$ & $1700\sim2600$ & $2500\sim3400$ \\
Recommended frequency range (MHz) & $500\sim950$ & $950\sim1750$ & $1750\sim2550$ & $2550\sim3300$ \\
Central frequency (MHz)  & 550 & 1350 & 2150 & 2950 \\
Local oscillator frequency (MHz)  & None & 1900 & 2700 & 3500 \\
Beam width HPBW (arcmin)           & $7.58\sim4.02$ & $4.42\sim2.29$ & $2.29\sim1.71$ & $1.71\sim1.57$ \\
High noise temperature $T_{\rm cal,\,\texttt{XX}}$ (K) & $17.1 \sim 21.6 $ & $15.1 \sim 17.0 $ & $11.6 \sim 15.1 $ & $13.0 \sim 14.9 $ \\
High noise temperature $T_{\rm cal,\,\texttt{YY}}$ (K) & $19.5 \sim 24.6 $ & $17.1 \sim 19.0 $ & $13.3 \sim 16.2 $ & $12.5 \sim 15.7 $ \\
Low noise temperature $T_{\rm cal,\,\texttt{XX}}$ (K) & $1.5 \sim 2.1 $ & $1.4 \sim 1.8 $ & $1.2 \sim 1.8 $ & $1.1 \sim 1.6 $ \\
Low noise temperature $T_{\rm cal,\,\texttt{YY}}$ (K) & $1.9 \sim 2.4 $ & $1.5 \sim 1.9 $ & $1.2 \sim 1.7 $ & $1.3 \sim 1.7 $ \\
Antenna gain (K/Jy) & $11.3 \sim 14.3 $ & $12.0 \sim 12.5 $ & $9.7 \sim 13.1 $ & $7.7 \sim 9.9 $ \\
Aperture efficiency $\eta$  & $0.44 \sim 0.56 $ & $0.47 \sim 0.49 $ & $0.38 \sim 0.51 $ & $0.30 \sim 0.38 $ \\
System temperature $T_{\rm sys}$ (K) & $88.9 \sim 112.3 $ & $96.8 \sim 104.1 $ & $100.0 \sim 114.7 $ & $120.9 \sim 130.5 $ \\
\hline
\end{tabular}}
\begin{flushleft}
\normalsize
\textbf{Notes.} Detailed parameter setups for the UWB receiver are presented in \citet{Liu2022}, e.g., the cryogenic microwave unit, the warm microwave, and the frequency mixing unit.
\end{flushleft}
\end{table*}

\section{Measurement Parameters of the UWB receiver}
\label{sec:uwb_para}

\subsection{The Noise Source}
\label{sec:noise_dipole}

\begin{figure*}
\centering
\includegraphics[width=0.90\textwidth, angle=0]{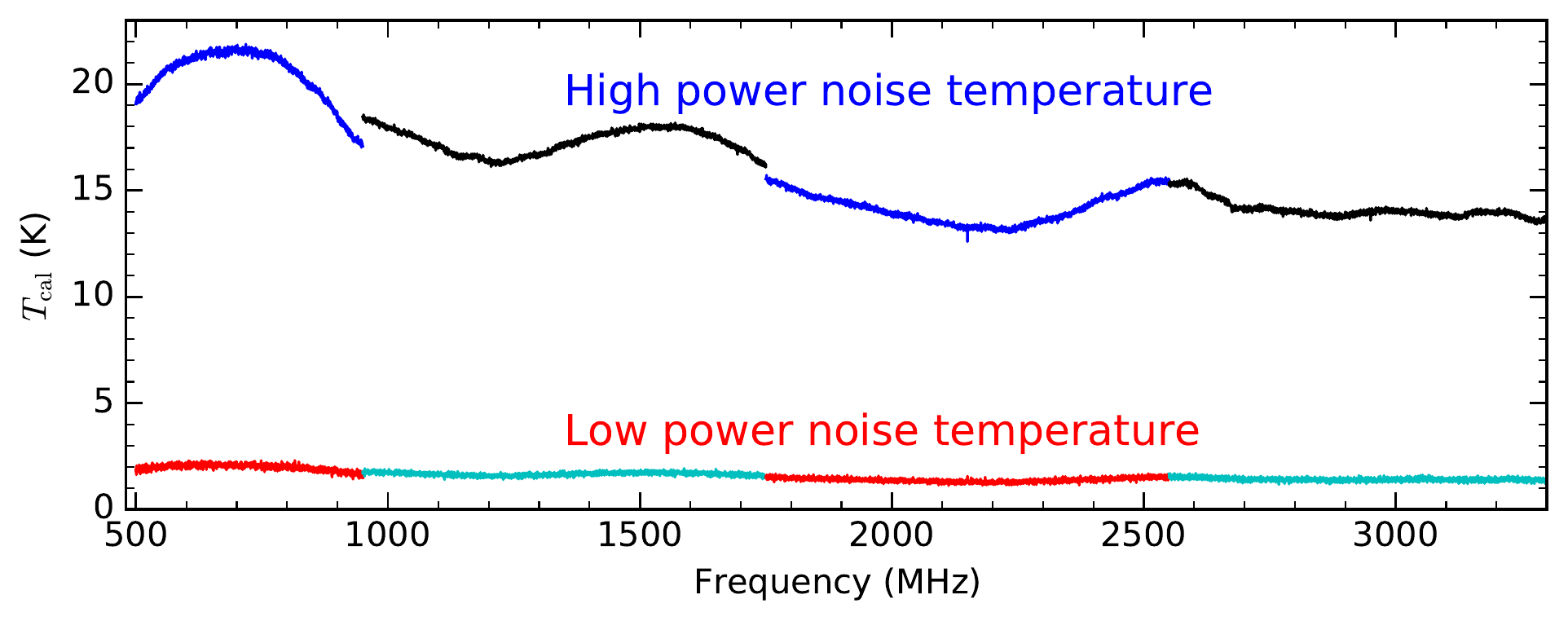}
\caption{The high and low power noise diode temperatures of the average of two polarizations \texttt{XX} and \texttt{YY} for the FAST UWB 500-3300\,MHz measured on June 17, 2022.}
\label{Fig:tcal}
\end{figure*}

Like FAST 19-beam array, the UWB receiver also contains a stabilized noise injection system \citep{Jiang2020}. The noise is injected between the feed and the low noise amplifiers. The noise source is a single diode whose signal is split into each polarization. The noise diode has two adjustable power output modes with 1.5-2.0\,K for low power noise temperature, and 13.5-22.0\,K for high power noise temperature. Based on testing a series of hot load measurements, the noise diode is stable and meet the requirements of data calibration. The low and high power noise temperatures are shown in Figure\,\ref{Fig:tcal} and listed in Table\,\ref{tab_uwb_gain_Tsys}. The full noise diode data for UWB 500-3300\,MHz could be download online.

\subsection{Beam Size}
\label{sec:beam}

\begin{figure*}
\centering
\includegraphics[width=0.90\textwidth, angle=0]{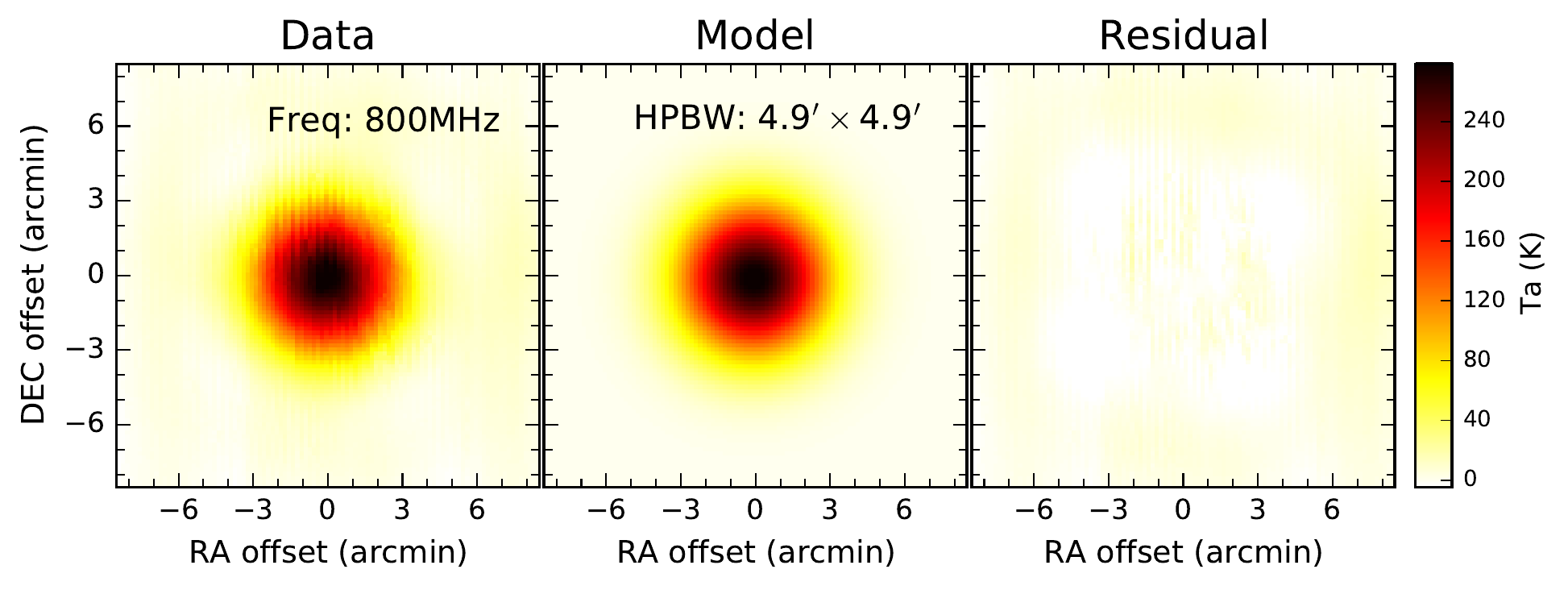}
\includegraphics[width=0.90\textwidth, angle=0]{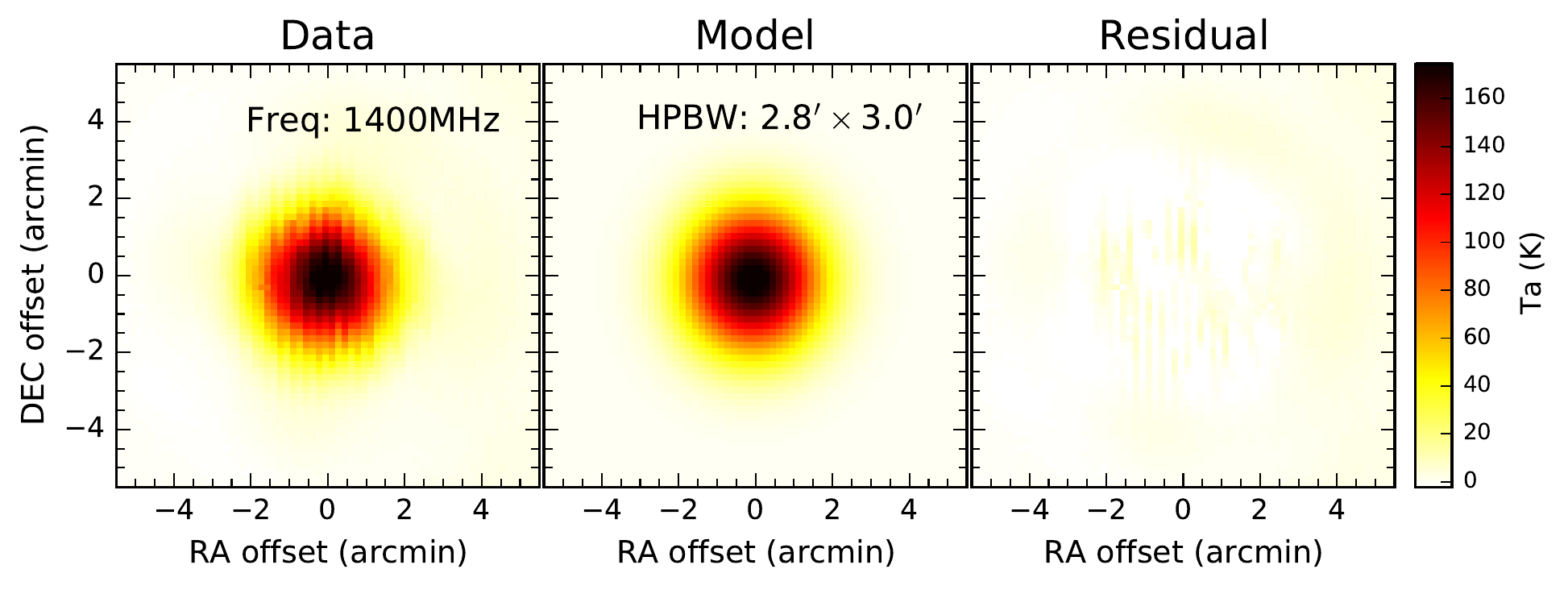}
\includegraphics[width=0.90\textwidth, angle=0]{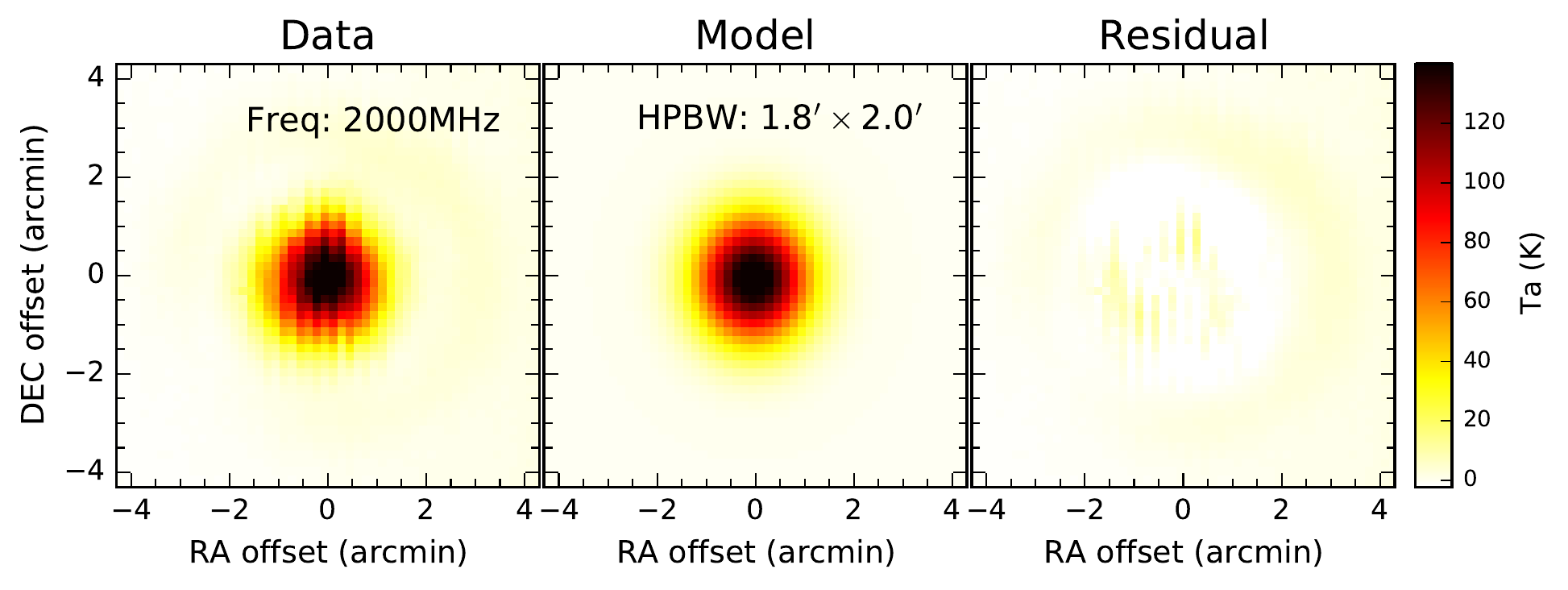}
\includegraphics[width=0.90\textwidth, angle=0]{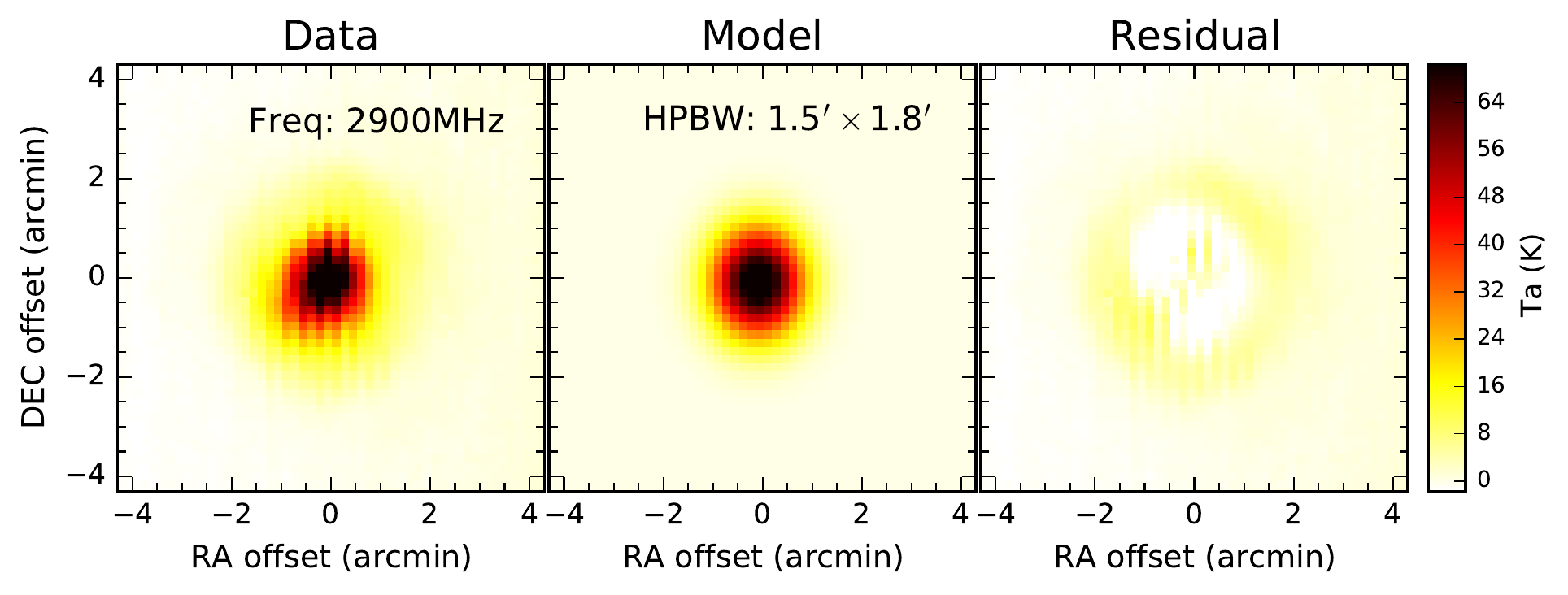}
\caption{The beam structures at 800, 1400, 2000, and 2900\,MHz measured by observing calibrator 3C286 on March 26, 2023. The used observation mode is OTF along the direction of the right ascension, and sampling time is 0.2s, scanning velocity is 20\,arcsec per second, and the scanning space is 12\,arcsec. }
\label{Fig:beam}
\end{figure*}

\begin{figure*}
\centering
\includegraphics[width=0.90\textwidth, angle=0]{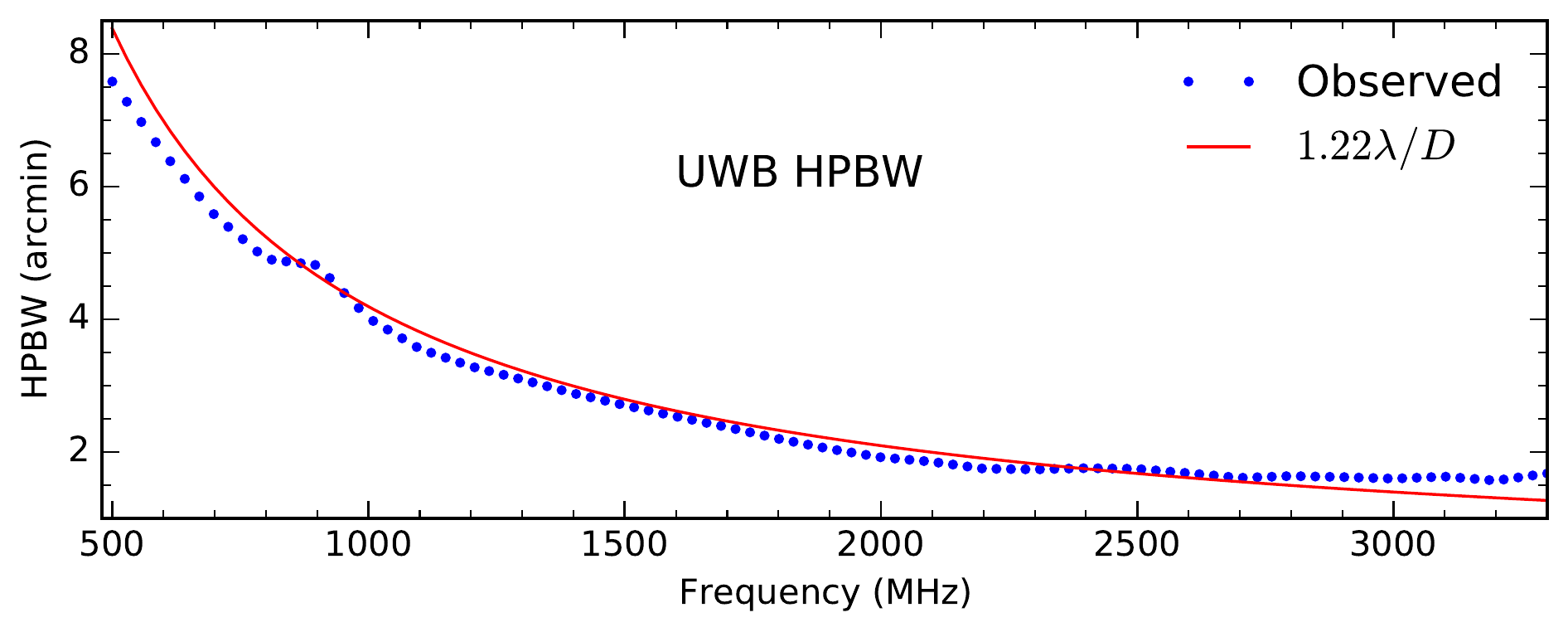}
\caption{The HPBW distribution (blue dotted line) for UWB 500-3300\,MHz measured by observing radio point source 3C286 within ZA of 26.4$^\circ$ on March 26, 2023. The red curve indicates the theoretical HPBW = $1.22\lambda/D$ with an assumed telescope diameter $D=300$\,m. }
\label{Fig:HPBW}
\end{figure*}

To measure the beam properties of FAST UWB receiver, we directly make mapping observations toward a radio point source 3C286 on the sky on March 26, 2023. The used observation mode is OTF along the direction of the right ascension, and sampling time is 0.2s, scanning velocity is 20\,arcsec per second, and the scanning space is 12\,arcsec. The mapping area is around $20'\times20'$, which is large enough for covering the whole beam structure at 500-3300\,MHz. Figure\,\ref{Fig:beam} displays examples of observed and fitted beam structures at 800, 1400, 2000, and 2900\,MHz. Table\,\ref{Fig:beam} lists all the measured HPBW at 500-3300\,MHz. Figure\,\ref{Fig:HPBW} shows the observed HPBW and the theoretical HPBW = $1.22\lambda/D$ with an assumed telescope diameter $D=300$\,m at 500-3300\,MHz. We find that below $\sim$2400\,MHz, the observed HPBW is smaller than the theoretical HPBW. This indicates that the telescope effective aperture is larger than 300\,m below $\sim$2400\,MHz. We noticed that the measured UWB HPBWs are consistent with the FAST 19-beam receiver between 1000-1500\,MHz.

\subsection{Pointing Accuracy}
\label{sec:pointing}

In FAST feed cabin, the UWB receiver has been placed at the phase center based on many pointing tests. According to antenna measurements, the UWB observations have the same pointing accuracy as FAST 19-beam array. The measured standard deviation of pointing accuracy is better than $\sim$7.9\,arcsec within zenith angle (ZA) of 26.4$^\circ$ \citep{Jiang2020}. For example, the measured pointing error is $\sim$7.0\,arcsec when measuring the beam structures using the radio point source 3C286 on March 26, 2023. The pointing accuracy of $\sim$7.0\,arcsec only takes around one twelfth of the HPBW (HPBW$_{\rm 3300\,MHz} \approx 1.6'$) at the frequency of 3300\,MHz for the FAST. Therefore, the pointing accuracy well meets the requirement of current UWB receiver observation.

\subsection{Antenna Gain and Aperture Efficiency}
\label{sec:gain}

\begin{figure*}
\centering
\includegraphics[width=0.90\textwidth, angle=0]{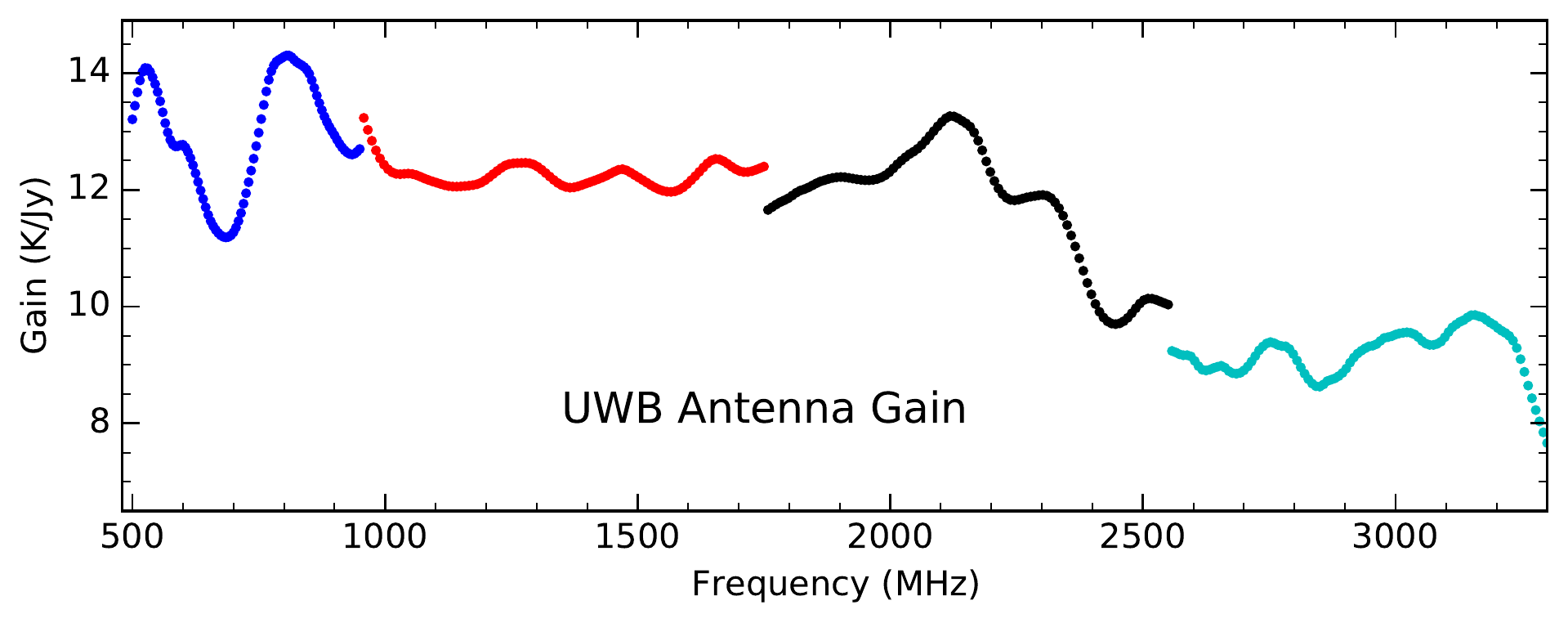}
\caption{The antenna gains within ZA of 26.4$^\circ$ for UWB 500-3300\,MHz measured by observing flux calibrator 3C286 on March 5, 2023. Four separated UWB bands are indicated with different colors. }
\label{Fig:gain}
\end{figure*}

Figure\,\ref{Fig:gain} shows the antenna gain distribution within ZA of 26.4$^\circ$ for UWB 500-3300\,MHz measured by observing a stable flux calibrator 3C286 on March 5, 2023. With absolute measurement of noise dipole, the UWB observed ON-OFF data could be calibrated to antenna temperature ($T_{\rm a, 3C286}$) in Kelvin. The flux density (in Jy) of 3C286 within UWB band could be fitted with a polynomial function \citep{Perley2017}
\begin{equation}
  {\rm log}(S_{\rm 3C286}) = 1.2481-0.4507\times[{\rm log}(\nu)]-0.1798\times[{\rm log}(\nu)]^2+0.0357\times[{\rm log}(\nu)]^3,
  \label{equ_3c286}
\end{equation}
where $S_{\rm 3C286}$ and $\nu$ are the flux density in Jy and the frequency in GHz, respectively. Then the antenna gain could be estimated as
\begin{equation}
  {\rm Gain} = \frac{T_{\rm a, 3C286}}{S_{\rm 3C286}}.
  \label{equ_gain}
\end{equation}
The derived UWB gain at $\sim$1400\,MHz is $\sim$12.0\,K\,Jy$^{-1}$, which is lower than that of the FAST 19-beam array ($\sim$16.0\,K\,Jy$^{-1}$), mainly because the UWB receiver is uncooled. Up to 3200\,MHz, the UWB gain is $\sim$9.5\,K\,Jy$^{-1}$. This meets the requirement of CH observation at $\sim$3263.794\,MHz. The full antenna gain parameters for UWB 500-3300\,MHz could be download online, and are partly listed in Table\,\ref{tab_uwb_gain_Tsys}.

Seen from Figure\,\ref{Fig:gain}, the antenna gain turns to be so low at high frequency end, probably because the reflector precision or the reflection efficiency turns to be low at such high frequency band. The wild fluctuation at low frequency end should be resulted from the serious RFI pollution at 500-920\,MHz. Generally, the variation of the monitored antenna gain is less than $\sim$10\% from August 2022 to March 2023. This indicates that the FAST UWB receiver is relatively stable, but it still needs long-time monitoring for better data calibration.

Assuming the efficiency aperture of FAST is 300\,m at 500-3300\,MHz, the corresponding geometric illumination area produces a theoretical gain with $G_0$ = 25.6\,K\,Jy$^{-1}$ \citep{Jiang2020}. The aperture efficiency $\eta$ of the FAST UWB receiver could be estimated by $\eta = {\rm Gain}/G_0$. The maximum and minimum gains are, respectively, 0.56 and 0.30 at 500-3300\,MHz. All derived aperture efficiencies are listed in Table\,\ref{tab_uwb_gain_Tsys}.

\subsection{System Temperature}
\label{sec:tsys}

\begin{figure*}
\centering
\includegraphics[width=0.90\textwidth, angle=0]{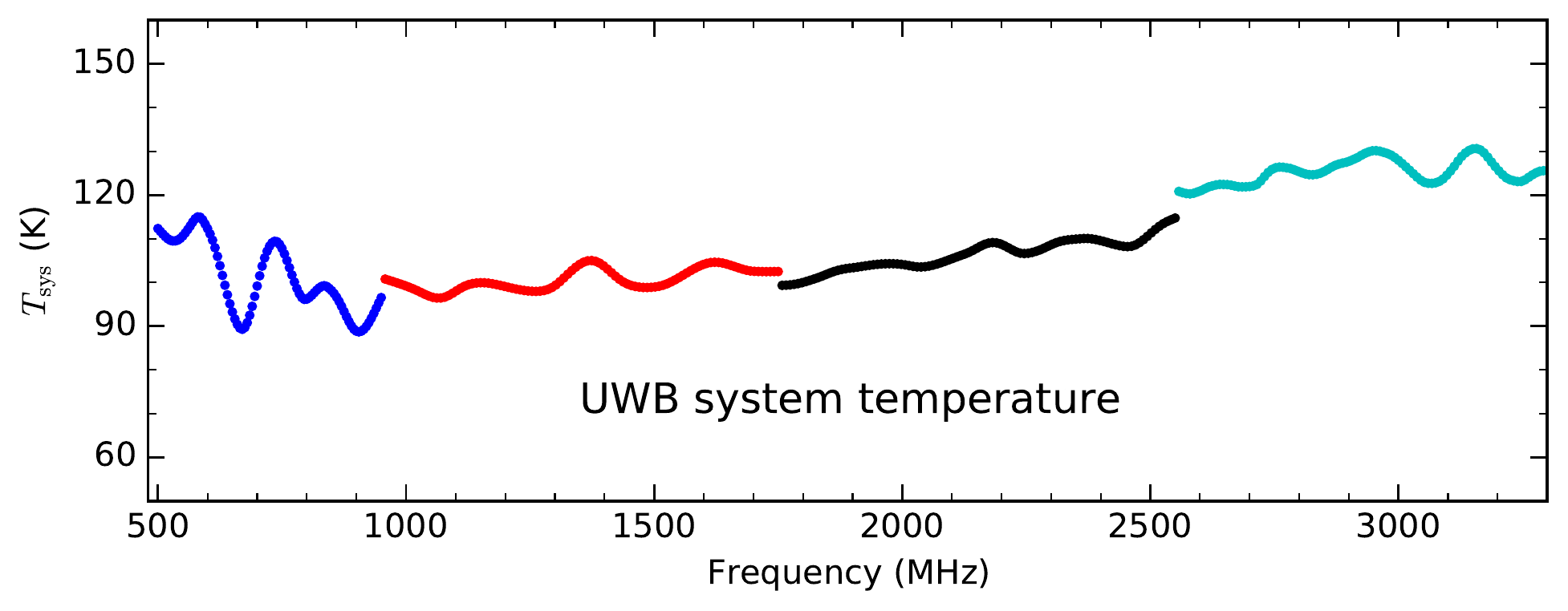}
\caption{The system temperature within ZA of 26.4$^\circ$ for UWB 500-3300\,MHz measured by observing cold sky on March 5, 2023. Four separated UWB bands are indicated with different colors. }
\label{Fig:Tsys}
\end{figure*}

System temperature is a synthetical contribution from noise of receiver ($T_{\rm rec}$), continuum brightness temperature of the sky ($T_{\rm sky}$), emission of the Earth’s atmosphere ($T_{\rm atm}$), and radiation of the surrounding terrain ($T_{\rm scat}$) \citep{Campbell2002,Jiang2020} as
\begin{equation}
  T_{\rm sys} = T_{\rm rec} + T_{\rm sky} + T_{\rm atm} + T_{\rm scat}.
  \label{equ_Tsys}
\end{equation}
Figure\,\ref{Fig:Tsys} displays the system temperature ($T_{\rm sys}$) within ZA of 26.4$^\circ$ for UWB 500-3300\,MHz measured by observing cold sky on March 5, 2023. The raw data were converted to antenna temperature with the noise data in Figure\,\ref{Fig:tcal}. The data points, which are deviated from the main curve, are resulted from strong RFI. The UWB system temperatures are 90-130\,K for the band of 500-3300\.MHz. The high system temperature mostly arises from the uncooled UWB receiver, whose parameters are presented in \citet{Liu2022}. The measured system temperature ($T_{\rm sys}$) for UWB 500-3300\,MHz are also listed in Table\,\ref{tab_uwb_gain_Tsys}. Such high system temperatures require long-time integration to compensate. In the future, once there is enough space in the feed cabin, the UWB receiver with cryogenic low-noise front-end will be installed, then the lower system noise temperature and higher detection sensitivity could be achieved.

\begin{table*}
\caption{Detailed parameters of noise diode temperature ($T_{\rm cal}$), antenna gain, aperture efficiency ($\eta$), system temperature ($T_{\rm sys}$), and beam width (HPBW) for UWB receiver.}
\label{tab_uwb_gain_Tsys} 
\centering \small  
\setlength{\tabcolsep}{1.6mm}{
\begin{tabular}{l|cccccccc}
\hline \hline
 Frequency & \multicolumn{2}{c}{High $T_{\rm cal}$} & \multicolumn{2}{c}{Low $T_{\rm cal}$} & Gain & $\eta$ & $T_{\rm sys}$ & HPBW \\
  & \texttt{XX} & \texttt{YY}  & \texttt{XX} & \texttt{YY} & & & Beam  \\
 MHz & K & K  & K & K & K & & K\,Jy$^{-1}$ & arcmin \\
\hline
  500 &  19.07 &  21.78 &   1.85 &   2.12 &  13.21 &   0.52 & 112.33 &   7.58 \\
  550 &  20.52 &  23.39 &   2.01 &   2.19 &  13.67 &   0.53 & 110.62 &   7.05 \\
  600 &  21.11 &  24.00 &   2.01 &   2.41 &  12.77 &   0.50 & 112.31 &   6.51 \\
  650 &  21.50 &  24.62 &   2.06 &   2.29 &  11.57 &   0.45 &  93.26 &   6.04 \\
  700 &  21.55 &  24.64 &   2.13 &   2.43 &  11.27 &   0.44 &  99.18 &   5.57 \\
  750 &  21.33 &  24.27 &   2.06 &   2.33 &  12.98 &   0.51 & 107.79 &   5.24 \\
  800 &  20.85 &  23.67 &   1.98 &   2.32 &  14.28 &   0.56 &  96.19 &   4.91 \\
  850 &  19.97 &  22.21 &   1.93 &   2.04 &  13.99 &   0.55 &  98.13 &   4.86 \\
  900 &  18.39 &  20.80 &   1.69 &   2.05 &  12.94 &   0.51 &  88.86 &   4.82 \\
  950 &  17.12 &  19.47 &   1.53 &   1.95 &  12.70 &   0.50 &  96.54 &   4.42 \\
 1000 &  16.92 &  18.96 &   1.61 &   1.81 &  12.41 &   0.48 &  99.19 &   4.02 \\
 1050 &  16.64 &  18.49 &   1.64 &   1.72 &  12.27 &   0.48 &  96.75 &   3.79 \\
 1100 &  16.07 &  17.92 &   1.60 &   1.78 &  12.13 &   0.47 &  97.98 &   3.55 \\
 1150 &  15.68 &  17.45 &   1.52 &   1.55 &  12.06 &   0.47 &  99.94 &   3.42 \\
 1200 &  15.52 &  17.15 &   1.42 &   1.62 &  12.18 &   0.48 &  99.06 &   3.29 \\
 1250 &  15.64 &  17.19 &   1.51 &   1.61 &  12.45 &   0.49 &  98.02 &   3.19 \\
 1300 &  15.97 &  17.42 &   1.65 &   1.67 &  12.40 &   0.48 &  99.28 &   3.09 \\
 1350 &  16.21 &  18.29 &   1.73 &   1.91 &  12.08 &   0.47 & 104.14 &   2.99 \\
 1400 &  16.37 &  18.49 &   1.52 &   1.74 &  12.11 &   0.47 & 103.62 &   2.88 \\
 1450 &  16.65 &  18.85 &   1.59 &   1.90 &  12.29 &   0.48 &  99.48 &   2.79 \\
 1500 &  16.82 &  19.00 &   1.62 &   1.75 &  12.22 &   0.48 &  98.96 &   2.70 \\
 1550 &  17.05 &  18.94 &   1.67 &   1.92 &  11.98 &   0.47 & 101.10 &   2.62 \\
 1600 &  16.96 &  18.95 &   1.57 &   1.83 &  12.11 &   0.47 & 104.14 &   2.54 \\
 1650 &  16.43 &  18.67 &   1.77 &   1.78 &  12.52 &   0.49 & 104.08 &   2.46 \\
 1700 &  15.89 &  18.03 &   1.47 &   1.69 &  12.33 &   0.48 & 102.57 &   2.37 \\
 1750 &  15.13 &  17.21 &   1.39 &   1.69 &  12.40 &   0.48 & 102.50 &   2.29 \\
 1800 &  13.95 &  16.22 &   1.43 &   1.68 &  11.87 &   0.46 & 100.01 &   2.20 \\
 1850 &  13.87 &  15.69 &   1.36 &   1.58 &  12.09 &   0.47 & 101.98 &   2.12 \\
 1900 &  13.61 &  15.24 &   1.24 &   1.48 &  12.22 &   0.48 & 103.36 &   2.05 \\
 1950 &  13.35 &  14.89 &   1.29 &   1.47 &  12.16 &   0.48 & 104.15 &   1.98 \\
 2000 &  13.33 &  14.53 &   1.23 &   1.44 &  12.32 &   0.48 & 104.12 &   1.92 \\
 2050 &  13.08 &  14.26 &   1.26 &   1.35 &  12.68 &   0.50 & 103.68 &   1.89 \\
 2100 &  13.25 &  13.78 &   1.38 &   1.45 &  13.14 &   0.51 & 105.42 &   1.85 \\
 2150 &  11.58 &  13.60 &   1.83 &   1.30 &  13.14 &   0.51 & 107.80 &   1.80 \\
 2200 &  12.90 &  13.32 &   1.23 &   1.31 &  12.27 &   0.48 & 108.70 &   1.75 \\
 2250 &  12.89 &  13.41 &   1.24 &   1.24 &  11.82 &   0.46 & 106.65 &   1.74 \\
 2300 &  13.30 &  14.00 &   1.29 &   1.38 &  11.91 &   0.47 & 108.61 &   1.74 \\
 2350 &  13.57 &  14.14 &   1.36 &   1.42 &  11.40 &   0.45 & 109.91 &   1.75 \\
 2400 &  14.42 &  14.66 &   1.33 &   1.48 &  10.17 &   0.40 & 109.56 &   1.76 \\
 2450 &  14.44 &  14.81 &   1.46 &   1.59 &   9.71 &   0.38 & 108.23 &   1.75 \\
 2500 &  14.89 &  15.54 &   1.45 &   1.48 &  10.10 &   0.39 & 111.14 &   1.75 \\
 2550 &  15.12 &  15.79 &   1.50 &   1.57 &  10.03 &   0.39 & 114.73 &   1.71 \\
 2600 &  14.90 &  15.71 &   1.38 &   1.52 &   9.10 &   0.36 & 120.91 &   1.68 \\
 2650 &  14.02 &  15.20 &   1.60 &   1.49 &   8.98 &   0.35 & 122.40 &   1.65 \\
 2700 &  13.39 &  14.59 &   1.32 &   1.66 &   8.91 &   0.35 & 121.93 &   1.61 \\
 2750 &  13.53 &  14.81 &   1.32 &   1.51 &   9.38 &   0.37 & 126.09 &   1.62 \\
 2800 &  13.45 &  14.68 &   1.50 &   1.43 &   9.15 &   0.36 & 125.32 &   1.64 \\
 2850 &  13.19 &  14.53 &   1.24 &   1.49 &   8.63 &   0.34 & 125.36 &   1.63 \\
 2900 &  12.97 &  14.68 &   1.32 &   1.37 &   8.91 &   0.35 & 127.69 &   1.62 \\
 2950 &  14.65 &  12.54 &   1.47 &   1.52 &   9.32 &   0.36 & 130.12 &   1.61 \\
 3000 &  13.37 &  14.71 &   1.12 &   1.64 &   9.52 &   0.37 & 127.93 &   1.60 \\
 3050 &  13.48 &  14.41 &   1.42 &   1.54 &   9.43 &   0.37 & 123.05 &   1.61 \\
 3100 &  13.50 &  14.09 &   1.31 &   1.45 &   9.50 &   0.37 & 124.84 &   1.63 \\
 3150 &  13.69 &  14.15 &   1.44 &   1.48 &   9.85 &   0.38 & 130.55 &   1.60 \\
 3200 &  13.72 &  14.22 &   1.46 &   1.38 &   9.65 &   0.38 & 125.84 &   1.57 \\
 3250 &  13.61 &  14.12 &   1.41 &   1.33 &   9.03 &   0.35 & 123.22 &   1.62 \\
 3300 &  13.51 &  13.77 &   1.50 &   1.35 &   7.66 &   0.30 & 125.56 &   1.68 \\
 \hline
\end{tabular}}
\begin{flushleft}
\end{flushleft}
\end{table*}

\section{Spectral-line backend and observations}
\label{sec:backend_obs}

\subsection{Backend}
\label{sec:backend}

At the backend, the whole UWB passband is separated into four subbands, 0-1100\,MHz, 800-1900\,MHz, 1600-2700\,MHz, and 2400-3500\,MHz. and each subband has 1048576 channels, so the frequency resolution is $\sim$1049.04\,Hz (or $\sim$1\,kHz). Any two adjacent bands have some overlapping frequency ranges to compensate for the shortcomings of the analog filter. The effective frequency ranges are 500-1000\,MHz, 900-1800\,MHz, 1700-2600\,MHz, and 2500-3400\,MHz, but the recommended frequency ranges for science observations are 500-950\,MHz for UWB-1, 950-1750\,MHz for UWB-2, 1750-2550\,MHz for UWB-3, and 2550-3300\,MHz for UWB-4 (see details in Table\,\ref{tab_uwb_para}). Combining the four subbands, the UWB could simultaneously and effectively cover the frequency ranging from 500 to 3300\,MHz (see Figure\,\ref{Fig:tcal}). The observed data are recorded in the spectral-line backend using a dual linear polarization (XX and YY) mode. Sampling time is adjustable, e.g., in 0.1s, 0.2s, 0.5s or 1.0s.

\subsection{Observation Modes}
\label{sec:obs_mode}

All the observation modes available in FAST 19-beam array could be used in the UWB receiver, such as Drift, OnOff, OTF, and so on \citep[see details in][]{Jiang2020}. However, we have to remember that the UWB has only one receiver available for observation. The setup parameters for scanning velocity is also the same as the FAST 19-beam array. The maximum scanning velocity is 15 and 30\,arcsec per second in direction of DEC and RA, respectively.

\subsection{Radio Frequency Interference}
\label{sec:rfi}

\begin{figure*}
\centering
\includegraphics[width=0.90\textwidth, angle=0]{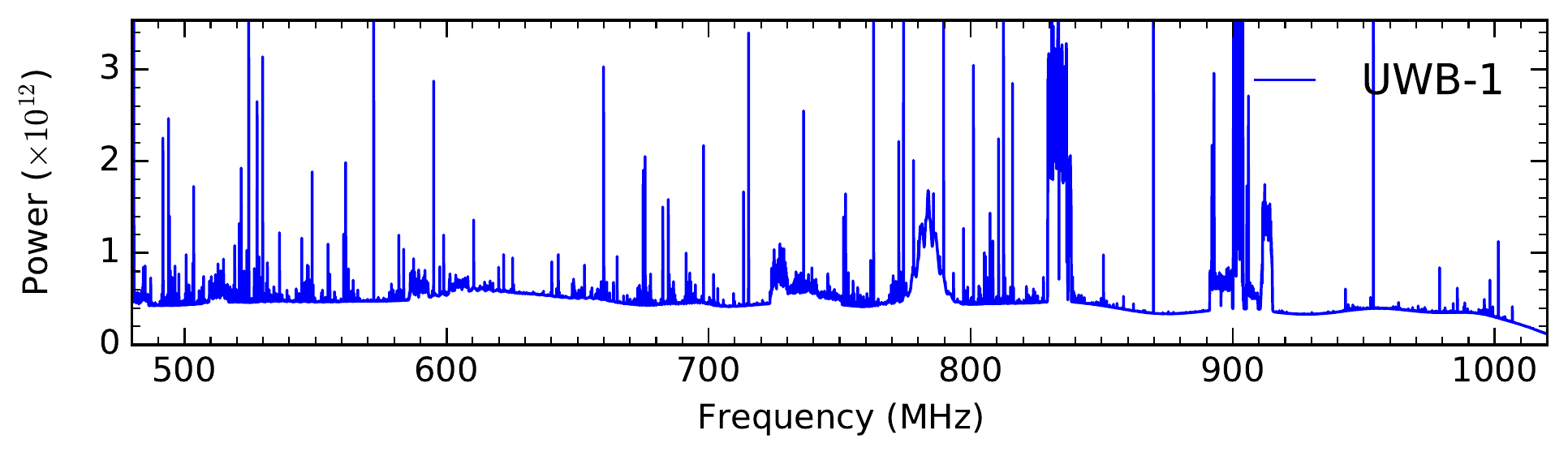}
\includegraphics[width=0.90\textwidth, angle=0]{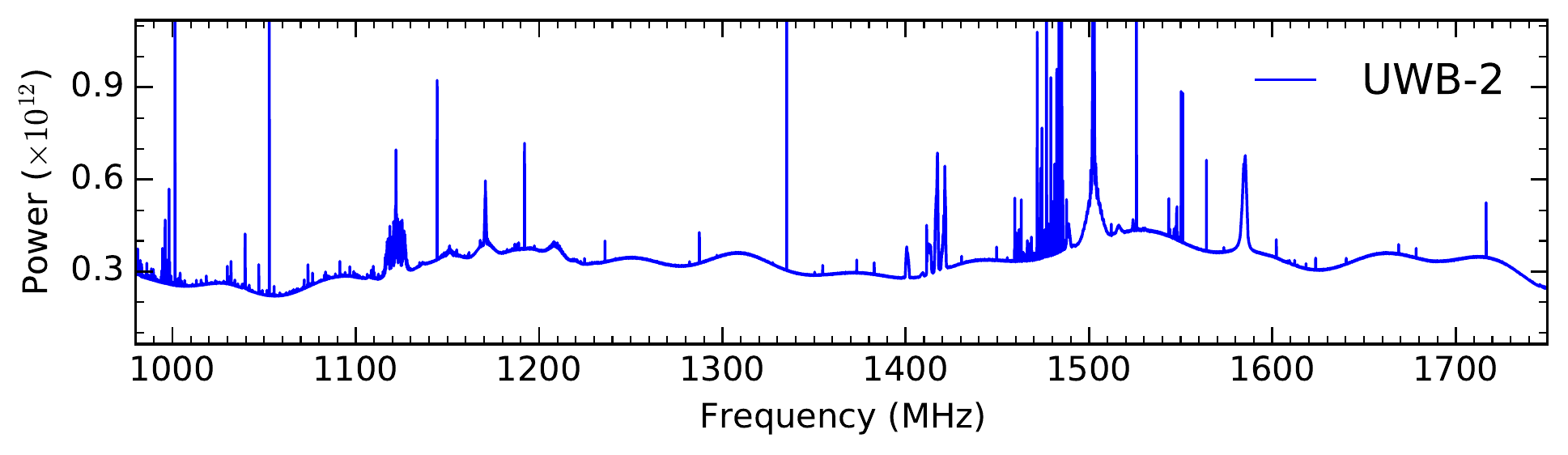}
\includegraphics[width=0.90\textwidth, angle=0]{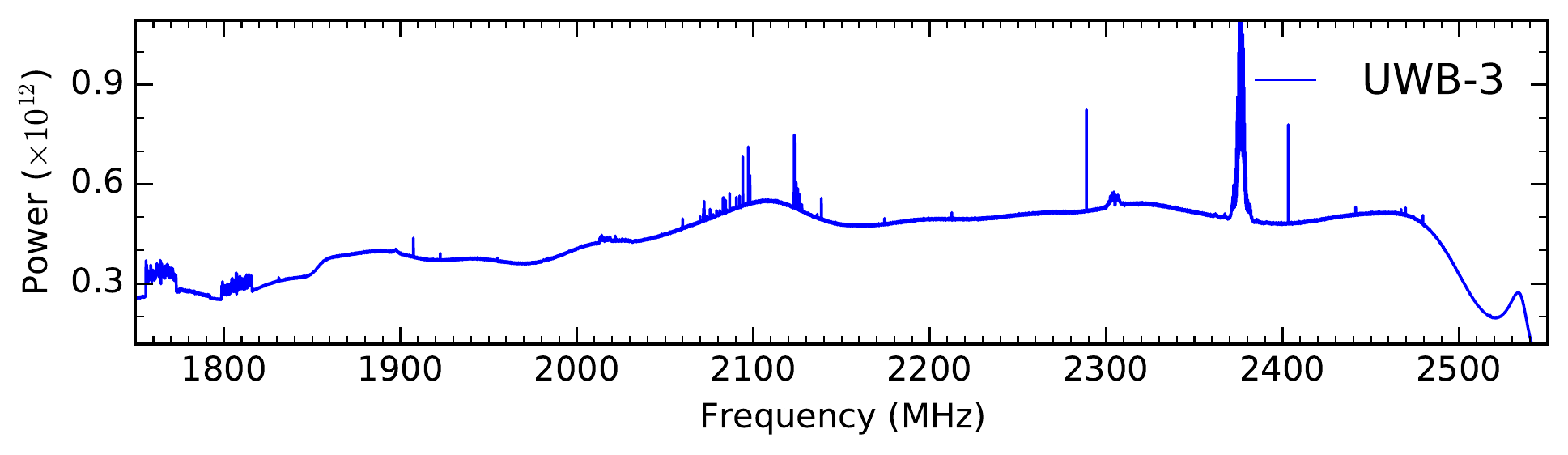}
\includegraphics[width=0.90\textwidth, angle=0]{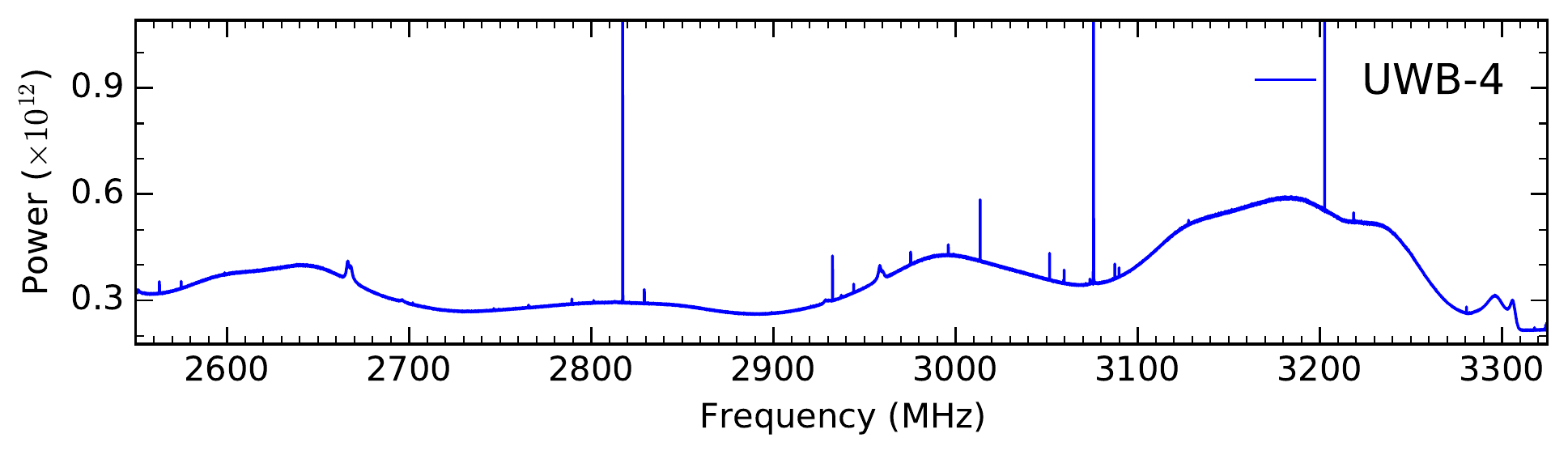}
\caption{The spectral bandpass and RFI distribution with one minute integration for UWB 500-3300\,MHz measured by observing cold sky on November 23, 2022. The emission lines basically are RFI, except \HI line at 1420\,MHz.}
\label{Fig:rfi}
\end{figure*}

In radio astronomy, radio frequency interference (RFI) becomes more and more serious for radio observational facilities \citep{Kesteven2005,An2017,Zeng2021,Zhang2022}. The RFI always influences the search and study of the interesting astronomical objects. Figure\,\ref{Fig:rfi} displays the whole bandwidth with one minute integration using UWB 500-3300\,MHz. In many tests, we found that in different sky directions, the RFI distribution at different frequencies is generally similar to that shown in Figure\,\ref{Fig:rfi}, but the intensities are varied. Additionally, the low frequency bands (500-950\,MHz) have a more serious RFI pollution than the other high frequency bands \citep{Zhang2020}. All the emission lines basically are RFI, except \HI line at 1420\,MHz. The extremely strong and evident RFI are from communication satellites and navigation satellites \citep{Wang2021}. Therefore, we must pay attention to avoiding the frequency positions of the strong RFI.

\subsection{\HI and OH lines}
\label{sec:lines}

\begin{figure*}
\centering
\includegraphics[width=0.70\textwidth, angle=0]{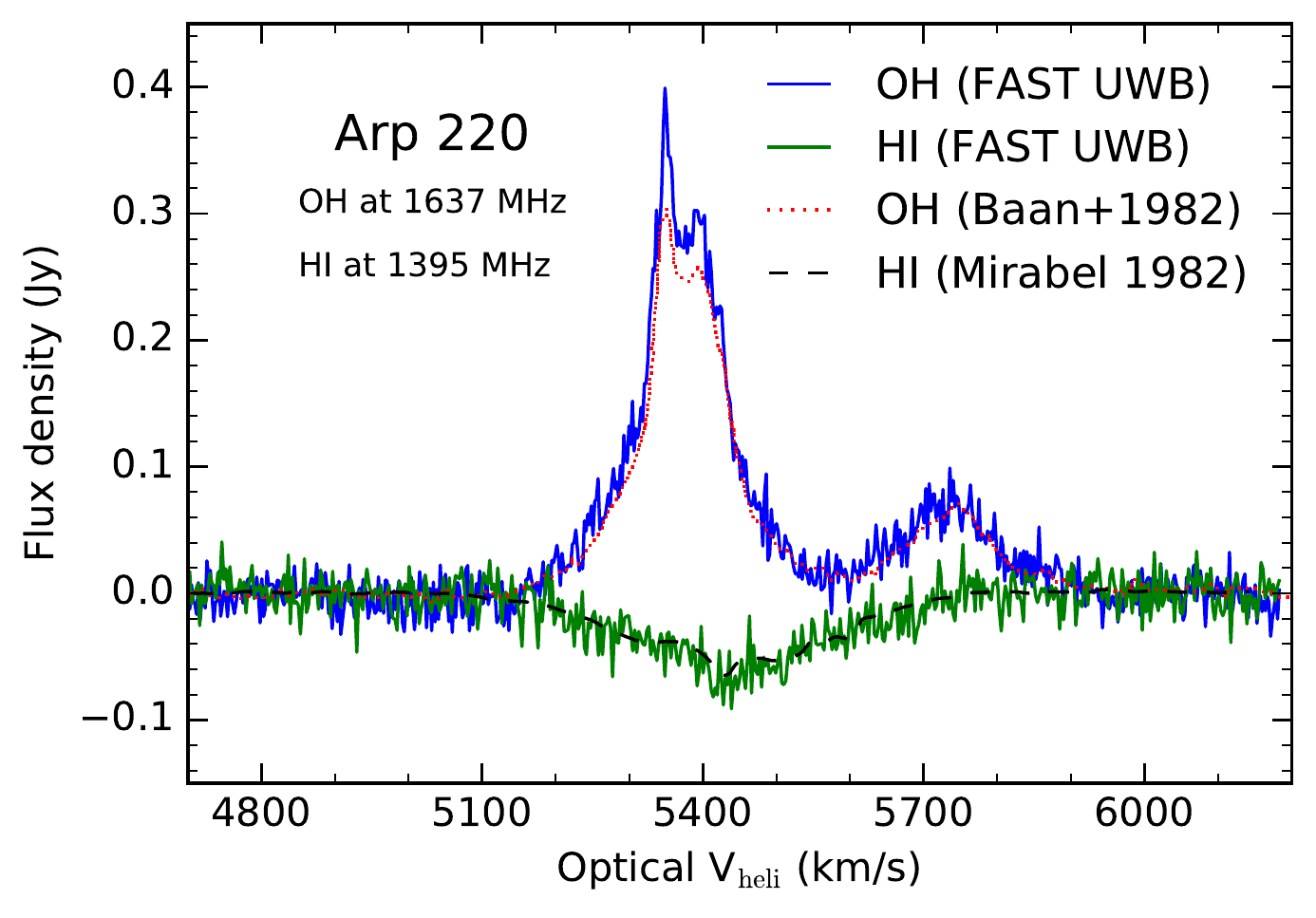}
\caption{Arp 220 (IC 4553) OH emission and \HI absorption lines simultaneously covered by UWB-3 and UWB-2 bands, respectively. Arp 220 is a well-known starburst Galaxy with redshift of 0.01840 \citep{Baan1982}. For the FAST UWB observed lines, the integration time is 10 minutes, and they have been smoothed into a frequency resolution of 12.0\,kHz, leading to an rms of 4.41\,mJy. The dotted and dashed curves present the OH and \HI lines observed by Arecibo 300\,m from \citet{Baan1982} and \citet{Mirabel1982}, respectively. The integration time is 25 minutes for the Arecibo OH line, but for the Arecibo \HI line there is no integration-time parameter recorded in \citet{Mirabel1982}.}
\label{Fig:arp220}
\end{figure*}

Figure\,\ref{Fig:arp220} shows Arp 220 (IC 4553) OH emission and \HI absorption lines observed by UWB receiver with 600s on-time integration. Arp 220 is a well-known starburst Galaxy with a redshift of 0.01840 \citep{Baan1982}. The observed redshift frequencies of the OH and \HI lines are, respectively, 1637 and 1395\,MHz, which are covered by UWB-4 and UWB-2 bands, respectively. In 600s integration, the measured spectral rms is around 15.27\,mJy with an original channel space of 1.0\,kHz. For the \HI absorption line of Arp 220 (see the \HI line in Figure\,\ref{Fig:arp220}), the measured flux density by the FAST UWB receiver is only $\sim$3\% higher than the Arecibo 300\,m observations \citep{Mirabel1982}. In addition, for the OH emission line of Arp 220 with the rest frequency of 1665.402\,MHz (see the right OH peak in Figure\,\ref{Fig:arp220}), the measured flux density by the UWB is also only $\sim$3\% higher than the Arecibo 300\,m observations \citep{Baan1982}. However, for OH emission line of Arp 220 at its rest frequency of 1667.359\,MHz (see the left OH peak in Figure\,\ref{Fig:arp220}), the measured flux density by the UWB is $\sim$10\% higher than the Arecibo 300\,m observations \citep{Baan1982}. This is probably because the OH flux density of Arp 220 at 1665.402\,MHz is variable \citep{Darling2002}. Generally, our measured (OH and \HI lines) flux density and velocity of Arp 220 are well consistent with Arecibo 300\,m observations \citep{Baan1982,Mirabel1982,Mirabel1988}. This further suggests that the FAST UWB receiver already has a good performance for spectral science observation at 500-3300\,MHz.

\section{Summary}
\label{sect:summary}

The Five-hundred-meter Aperture Spherical radio Telescope (FAST) has already been well running for several years since FAST began its commission when the construction was completed on September 25, 2016. The 19-beam receiver covering 1.05-1.45\,GHz was used for most of the science observations. However, high frequency observations, e.g., OH lines at rest frequencies of 1665 and 1667\,MHz, are needed to study the star formation in the Milky Way and nearby galaxies. The designed FAST reflector precision actually has met the observational requirement at the high frequency of around 3000\,MHz.

Fortunately, a new uncooled ultra wideband (UWB) receiver, simultaneously covering 500-3300\,MHz, has been mounted in the FAST feed cabin on June 2022, and has passed a series of observational tests recently. The whole UWB band has been separated into four independent bands, but the recommended frequency ranges for users are UWB-1 for 500-950\,MHz, UWB-2 for 950-1750\,MHz, UWB-3 for 1750-2550\,MHz, and UWB-4 for 2550-3300\,MHz. Each band has 1048576 channels in the total frequency range, resulted in a high enough spectral resolution of 1\,kHz. At 500-3300\,MHz, the antenna gain is around 14.3-7.7\,K\,Jy$^{-1}$, the aperture efficiency is around 0.56-0.30, the system temperature is around 88-130\,K, and the HPBW is around 7.6-1.5\,arcmin. The measured antenna parameters above have been listed in the Table\,\ref{tab_uwb_gain_Tsys} for data reduction. The measured standard deviation of pointing accuracy is better than $\sim$7.9\,arcsec, when zenith angle is within 26.4$^\circ$. In addition, the sensitivity and stability of the UWB receiver are confirmed to satisfy expectation by spectral \HI and OH observations. By comparison, the measured Arp 220 (OH and \HI lines) flux density and velocity are well consistent with Arecibo 300\,m observations. This further suggests that the FAST UWB receiver already has a good performance for taking sensitive observations in various scientific goals at 500-3300\,MHz.

In the future, once there is enough space in the FAST feed cabin, the UWB receiver with cryogenic low-noise front-end will be installed, then the performance of UWB receiver will be highly improved. For example, the system temperature would decrease $\sim$50\,K, and the antenna gain would increase $\sim$2.5\,K\,Jy$^{-1}$. That will help us to take more sensitive observations in more various scientific goals than its current status.

\section*{Acknowledgements}
\addcontentsline{toc}{section}{Acknowledgements}

This work is supported by the National Key R\&D Program of China No. 2018YFE0202900. C.P.Z acknowledges support by the West Light Foundation of the Chinese Academy of Sciences (CAS). C.C. and H.F.L thank support by the National Natural Science Foundation of China Nos. 11803044, 11933003, 12173045, and 12273072. This work is sponsored partly by the CAS South America Center for Astronomy (CASSACA) and the China Manned Space Project NO. CMS-CSST-2021-A05. FAST is a Chinese national mega-science facility, operated by the National Astronomical Observatories of CAS (NAOC). We also wish to thank the anonymous referee for comments and suggestions that improved the clarity of the paper.

\bibliographystyle{raa}
\bibliography{references}

\end{document}